\documentclass[floatfix,twocolumn,aps,prx,reprint,superscriptaddress]{revtex4-2}

\usepackage{physics}
\usepackage{tabularx}
\usepackage{amsmath}
\usepackage{amssymb}
\usepackage[normalem]{ulem}
\usepackage{float}
\usepackage{graphicx}
\usepackage{color,soul}
\usepackage[pdftex, colorlinks, allcolors=blue]{hyperref}
\usepackage{gensymb}
\usepackage[caption=false]{subfig}
\usepackage[colorinlistoftodos]{todonotes}
\usepackage{lipsum}
\usepackage[normalem]{ulem}

\usepackage{tikz}
\usepackage{circuitikz}
\usepackage{tikz-timing}
\usetikzlibrary{arrows.meta,decorations.pathmorphing,decorations.pathreplacing,positioning,shapes,fadings}

\newcommand{\siswap}{\sqrt{i\mathrm{SWAP}}}
\newcommand{\sbswap}{\sqrt{b\mathrm{SWAP}}}

\makeatletter
\@ifundefined{textcolor}{}
{%
 \definecolor{BLACK}{gray}{0}
 \definecolor{WHITE}{gray}{1}
 \definecolor{RED}{rgb}{1,0,0}
 \definecolor{GREEN}{rgb}{0,1,0}
 \definecolor{BLUE}{rgb}{0,0,1}
 \definecolor{CYAN}{cmyk}{1,0,0,0}
 \definecolor{MAGENTA}{cmyk}{0,1,0,0}
 \definecolor{YELLOW}{cmyk}{0,0,1,0}
}
\usepackage{dcolumn}
\usepackage{bm}
\usepackage{float}

\makeatother

\begin{document}

\title{
Tunable inductive coupler for high-fidelity gates between fluxonium qubits
}

\author{Helin Zhang}
\thanks{These authors contributed equally to this work}
\affiliation{James Franck Institute, University of Chicago, Chicago, Illinois 60637, USA}
\affiliation{Department of Physics, University of Chicago, Chicago, Illinois 60637, USA}

\author{Chunyang Ding}
\thanks{These authors contributed equally to this work}
\affiliation{James Franck Institute, University of Chicago, Chicago, Illinois 60637, USA}
\affiliation{Department of Physics, University of Chicago, Chicago, Illinois 60637, USA}
\affiliation{Department of Physics and Applied Physics, Stanford University, Stanford, California 94305}

\author{D. K. Weiss}
\affiliation{Department of Physics and Astronomy, Northwestern University, Evanston, Illinois 60208, USA}
\affiliation{Yale Quantum Institute, Yale University, New Haven, CT 06511, USA}

\author{Ziwen Huang}
\email{Current address: Superconducting Quantum Materials and Systems Center,
Fermi National Accelerator Laboratory (FNAL), Batavia, IL 60510, USA}
\affiliation{Department of Physics and Astronomy, Northwestern University, Evanston, Illinois 60208, USA}

\author{Yuwei Ma}
\affiliation{James Franck Institute, University of Chicago, Chicago, Illinois 60637, USA}
\affiliation{Department of Physics, University of Chicago, Chicago, Illinois 60637, USA}

\author{Charles Guinn}
\affiliation{Department of Physics, Princeton University, Princeton, New Jersey 08544, USA}

\author{Sara Sussman}
\affiliation{Department of Physics, Princeton University, Princeton, New Jersey 08544, USA}

\author{Sai Pavan Chitta}
\affiliation{Department of Physics and Astronomy, Northwestern University, Evanston, Illinois 60208, USA}

\author{Danyang Chen}
\affiliation{Department of Physics and Astronomy, Northwestern University, Evanston, Illinois 60208, USA}

\author{Andrew A. Houck}
\affiliation{Department of Physics, Princeton University, Princeton, New Jersey 08544, USA}

\author{Jens Koch}
\affiliation{Department of Physics and Astronomy, Northwestern University, Evanston, Illinois 60208, USA}

\author{David I. Schuster}
\email{Corresponding author: dschus@stanford.edu}
\affiliation{James Franck Institute, University of Chicago, Chicago, Illinois 60637, USA}
\affiliation{Department of Physics, University of Chicago, Chicago, Illinois 60637, USA}
\affiliation{Department of Physics and Applied Physics, Stanford University, Stanford, California 94305}
\affiliation{Pritzker School of Molecular Engineering, University of Chicago, Chicago, Illinois 60637, USA}


\begin{abstract}
The fluxonium qubit is a promising candidate for quantum computation due to its long coherence times and large anharmonicity. We present a tunable coupler that realizes strong inductive coupling between two heavy-fluxonium qubits, each with $\sim50$MHz frequencies and $\sim5$ GHz anharmonicities. The coupler enables the qubits to have a large tuning range of $\textit{XX}$ coupling strengths ($-35$ to $75$ MHz). The $\textit{ZZ}$ coupling strength is $<3$kHz across the entire coupler bias range, and $<100$Hz at the coupler off-position. These qualities lead to fast, high-fidelity single- and two-qubit gates. By driving at the difference frequency of the two qubits, we realize a $\siswap$ gate in $258$ns with fidelity $99.72\%$, and by driving at the sum frequency of the two qubits, we achieve a $\sbswap$ gate in $102$ns with fidelity $99.91\%$. This latter gate is only 5 qubit Larmor periods in length. We run cross-entropy benchmarking for over $20$ consecutive hours and measure stable gate fidelities, with $\sbswap$ drift ($2 \sigma$) $< 0.02\%$ and $\siswap$ drift $< 0.08\%$.
\end{abstract}

\maketitle

\section{Introduction}

\begin{figure*}[t]
\centering
\includegraphics[width=\textwidth]{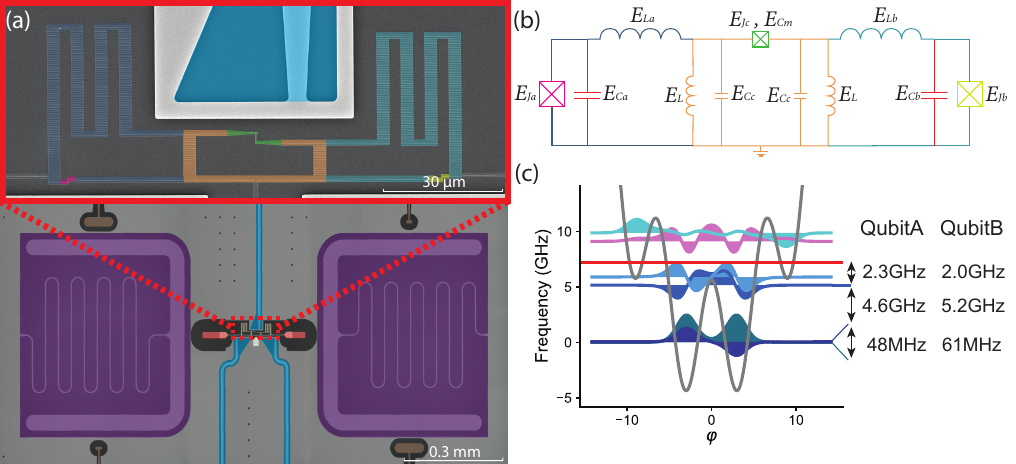}
\caption{Device, circuit and energy level diagram. (a) Top panel: False-colored scanning electron microscope image of the two fluxoniums and coupler junction loops. Each qubit consists of a small Josephson junction (pink and yellow) and an array of large junctions as an inductor (blue and cyan). The coupler consists of a small Josephson junction (green) and a shorter array of junctions (orange). Bottom panel: optical microscope image of the flux control lines (light blue), readout resonators (purple), qubit shunting capacitors (red), and resonator drive and readout lines (brown). (b) Circuit diagram for the coupled fluxoniums with the tunable coupler. The colors of the components correspond to the colors in the device images. (c) Energy-level diagram of the heavy-fluxonium wavefunctions at the flux-frustration point ($\Phi_{\rm ext} = \Phi_0/2$), plotted using qubit B parameters. The gray line represents the potential well, and the red line is at the readout resonator frequency. The first six energy eigenstate wavefunctions are plotted with solid colors. The qubit frequencies, anharmonicities, and resonator detunings for qubits A and B are indicated with black arrows.}
\label{fig:fig1}
\end{figure*}

Superconducting circuits are a promising platform for the development of scalable, error-corrected quantum computation, heralded by many recent advances towards large scale quantum processors~\cite{arute2019quantum, kim2023evidence} and continued improvements on performances of qubits and gate operations~\cite{devoret2013superconducting, krantz2019quantum, sivak2023real}. These advances have relied on the transmon qubit~\cite{koch2007}, and through collective effort, these qubits have achieved excellent coherence times~\cite{place2021new,sivak2023real} and gate fidelities approaching the quantum error correction threshold~\cite{Fowler2012Surface, Sung2021, Negrneac2021high}.

A promising alternative to the transmon is the fluxonium qubit~\cite{manucharyan2009fluxonium}, which has attractive properties including a nearly degenerate ground state and a large anharmonicity. Compared to other qubits fabricated with similar material qualities, low frequency qubits have less dielectric loss and thus a slower decoherence rate. The large anharmonicity mitigates the speed limit for qubit operations in transmons, and enables quantum gates as fast or even faster despite low qubit frequencies~\cite{Zhang2021Universal}. Recent works have demonstrated $\sim1$ms coherence times and $>99.9\%$ fidelity single-qubit gates using fluxonium qubits~\cite{nguyen2019high, somoroff2021millisecond, ding2023high}. In addition, one can flux-tune the fluxonium qubit to be noise biased, increasing energy coherence ($T_1 > 5$ms) at the expense of being more phase sensitive ~\cite{nate2018fluxonium, lin2018demonstration}.

Beyond single qubit operations, several recent works have demonstrated high-fidelity two-qubit fluxonium gates using either fixed capacitive coupling~\cite{ficheux2021fast, dogan2022demonstration, bao2022fluxonium, Xiong2022Arbitrary} or a tunable capacitive coupler~\cite{moskalenko2022high, ding2023high}. While these two-qubit gate schemes are promising, they either populate states outside of the computational subspace or use high-frequency fluxonium qubits. In this work, we realize a tunable \emph{inductive} coupler, and perform ($>99.9\%$) fidelity two-qubit gates. Our inductive coupler, similar to the g-mon coupler~\cite{chen2014qubit,Geller2015gmon}, realizes a large interaction strength even for low-frequency qubits, without involving any higher energy levels.  This enables the gate to take advantage of the full coherence of the fluxonium while avoiding any leakage to states outside the logical subspace.

In Section \ref{sec:coupling}, we provide theoretical analysis of our tunable inductive coupler, showing the origins of $\textit{XX}$ coupling. We also find that both the $\textit{XX}$ and unwanted $\textit{ZZ}$ coupling can be turned off, allowing for an operational spot for single qubit gates. In Section \ref{sec:device}, we report the characterization of our device near this operational spot, measuring the $\textit{ZZ}$ coupling to be $<100$ Hz. Finally, in Section \ref{sec:gates}, we construct $\siswap$ and $\sbswap$ gates using this $\textit{XX}$ interaction. We demonstrate the tuning of these gates and estimate their fidelities using various benchmarking protocols.

\section{Inductively coupled heavy-fluxonium qubits}
\label{sec:coupling}

In this letter, we use a tunable inductive coupler, inspired by designs from~\cite{chen2014qubit, Geller2015gmon, FQCE1, FQCE2}, to realize strong coupling between two fluxonium qubits. Each qubit is individually described by the Hamiltonian
\begin{equation}
   H_f = -4E_C\frac{d^2}{d\varphi^2}-E_J \cos(\varphi)+\frac{1}{2}E_L\left(\varphi + 2\pi \frac{\Phi_{\mathrm{ext}}}{\Phi_0}\right)^2,
\end{equation}
where $E_C=e^2/(2C_q)$ denotes the charging energy, $C_q$ the qubit's total shunting capacitance, $E_J$ the Josephson energy of the small junction, $E_L = \Phi_0^2/(4\pi^2L_{JA})$ the inductive energy and $L_{JA}$ the total inductance of the superinductor. $\Phi_{0}=h/2e$ is the superconducting flux quantum, and $\Phi_{\rm ext}$ denotes the flux threading the loop. We design heavy-fluxonium qubits [Fig.~\ref{fig:fig1}(a)] with large $E_{J}/E_{C}$, such that at half-integer flux bias, our qubits, labeled as $a, b$, have small splittings ($\omega_{a}/2\pi=48$ MHz, $\omega_{b}/2\pi=61$ MHz) and large anharmonicities ($\alpha_a=4.6$ GHz, $\alpha_b=5.2$ GHz). See Fig.~\ref{fig:fig1}(c) for the qubit level structure at the sweet spot ($\Phi_{\rm ext} = \Phi_0/2$) and Table~\ref{table:freqs} for the qubit parameters. 

These two heavy-fluxonium qubits are linked via an inductive coupler, as shown in Fig.~\ref{fig:fig1}(b). This full circuit, analyzed in Appendix~\ref{app:full_circuit}, consists of four degrees of freedom: the two fluxonium qubits and the two coupler degrees of freedom - one harmonic and one fluxonium-like. Ignoring the negligible cross-capacitance, these two fluxonium qubits do not directly interact with each other; instead, they each share an inductance ($E_L$) with the tunable coupler. Although the galvanic coupling is relatively strong, the coupler excitation energies ($9.5$GHz) are well above the energies associated with qubit excitations, leading to a dispersive interaction. Thus, near half-integer flux for the qubits, we can use a perturbative treatment, creating an effective Hamiltonian up to 4th order~\cite{Weiss2022}:

\begin{align}
\label{eq:effHam}
H_{\text{eff}} &= -\sum_{\mu=a,b}\frac{\omega_{\mu}}{2}\sigma_{z}^{\mu}-\Omega_{\mu}\sigma_{x}^{\mu} + J\sigma_{x}^{a}\sigma_{x}^{b}
+\zeta\sigma_{z}^{a}\sigma_{z}^{b},
\end{align}
where $\sigma_{x}^{\mu}, \sigma_{z}^{\mu}$
are the qubit Pauli operators in the basis of symmetric and anti-symmetric wavefunctions.

The first term in the effective Hamiltonian describes the qubit frequencies, including the Lamb shifts induced by the coupler. The second term captures the effect of qubit flux bias, which we use for single-qubit gates~\cite{Zhang2021Universal}. This term additionally incorporates a first-order perturbative shift due to the fluxonium-like coupler degree of freedom. The third term is the desired $\textit{XX}$ coupling, where $J$ is a function of the coupler flux $\Phi_{\mathrm{ext},c}$. Virtual exchanges through coupler excitations leads to a $\varphi_a \varphi_b$ term, which can be truncated to $\sigma_x^a \sigma_x^b$. There are two contributions here: a static one, due to the harmonic coupler degree of freedom, and a flux-tunable one, from the fluxonium-like coupler degree of freedom (see Ref. ~\cite{Weiss2022}). 
Because these contributions have opposite signs, we can null $J$ at a particular $\Phi_{\mathrm{ext}, c}$, which we term the ``off position.''. The final term describes a small unwanted $\textit{ZZ}$ interaction that comes from fourth-order perturbation theory.

This effective Hamiltonian governs our operation of the two-qubit device: a parametric drive controls $J$, allowing us to achieve two-qubit gates, and a particular $\Phi_{\mathrm{ext}, c}$ bias fully turns off the qubit-qubit coupling. However, since $\Omega_{\mu}$ is coupler-flux dependent, this tuning causes shifts in qubit frequencies. This can be compensated by an accompanying change in individual qubit fluxes ($\Phi_{\mathrm{ext},a}, \Phi_{\mathrm{ext},b}$) to hold each qubit at its sweet spot. This set of constraints defines a curve in the 3D space of flux parameters, which we term the ``sweet-spot contour''. 
Along the sweet-spot contour, $\zeta$ is generally non-zero, but we find numerically $\zeta<3$kHz. Effectively, this $\textit{ZZ}$ term can be cancelled via small ($< 10^{-5} \Phi_0$) flux shifts away from the qubit sweet spot. We find the flux noise at this bias point not to be the dominant dephasing mechanism, as it would only limit the pure qubit dephasing time to $1$ms, see Appendix~\ref{app:ZZ}. 

\begin{figure}[ht]
\centering
\includegraphics[width=\columnwidth]{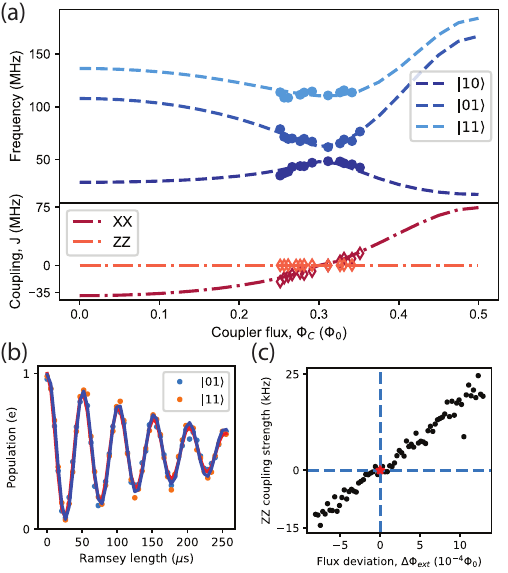}
\caption{
Measurements of $\textit{XX}$ and $\textit{ZZ}$ coupling strength. (a) Top panel: Measured $\ket{01}$, $\ket{10}$, and $\ket{11}$ frequencies as a function of coupler flux as the qubits are tuned to be along the ``sweet-spot contour''. The dashed lines are from full numerical simulations with parameters extracted from qubit spectroscopy, see Appendix~\ref{app:full_circuit}. Bottom panel: $\textit{XX}$ and $\textit{ZZ}$ coupling strength (diamonds) extracted from the data (dots). Numerical simulation predictions are plotted as the dashed lines. $\textit{ZZ}$ coupling strength is $<3$kHz across the entire coupler flux range. (b) Ramsey experiment of qubit B at the coupler ``off position'', with the qubits initialized in either $\ket{01}$ or $\ket{11}$. The frequency difference $(f_{11}-f_{10}) - (f_{01}-f_{00})$, extracted using Ramsey fits with different qubit initial states, measures the $\textit{ZZ}$ coupling strength to be $<100$Hz. (c) We perform this same measurement on qubit A while sweeping its flux away from the ``sweet-spot contour''. We find that $\textit{ZZ}$ coupling strength becomes non-zero, which we use as an indicator for fine-tuning qubit flux biases.}
\label{fig:fig2}
\end{figure}

\section{Device characterization}
\label{sec:device}

We measure the properties of a fabricated 2D superconducting device, validating our theoretical analysis of the tunable inductive coupler and calibrating it for single- and two-qubit gates. We use a tantalum base layer for increased qubit coherence~\cite{place2021new, wang2022towards}, and use double-angle aluminum evaporation~\cite{dolan1977offset} to fabricate the Josephson junctions (see Appendix ~\ref{app:fab}). The geometry of the device is shown in Figure~\ref{fig:fig1}(a). There are 205 Josephson junctions in each qubit's superinductor ($E_{La}, E_{Lb}$) and 17 junctions in each superinductor of the tunable coupler ($E_{L}$). There are three dedicated control lines, one for each of the qubits and one for the coupler, which are used for both DC biasing as well as RF flux driving (see Appendix~\ref{app:wiring}). 

There are two lumped LC resonators capacitively coupled to the fluxoniums for readout. These readout resonators have linewidths $\approx1.1$ MHz and dispersive shifts $\approx0.7$MHz from the computational levels. Since the qubit frequencies correspond to a temperature ($\sim3$mK) that is lower than the environment temperature, we must always initialize the qubit. We use a measurement-based active reset protocol for initialization~\cite{riste2012initialization, gebauer2020state}, enabled by the QICK controlled RFSoC FPGA~\cite{QICK}. The reset is carried out by measuring the qubit state and conditionally playing a single-qubit $\pi$ pulse, completed within $800$ns. With this method, we initialize the qubits with $\sim95\%$ fidelity, primarily limited by our measurement infidelity (which could be improved by using quantum-noise limited amplifiers, see Appendix~\ref{app:readout}).

To determine the coupler parameters, we scan the coupler flux over the range of $\Phi_{\mathrm{ext}, c}/\Phi_{0}=0.24$ to $\Phi_{\mathrm{ext}, c}/\Phi_{0}=0.34$ and measure qubit frequencies while staying on the ``sweet-spot contour'', shown in Fig.~\ref{fig:fig2}(a). Since $\Omega_{a,b}$ is zero along this contour, we determine the $\textit{XX}$ coupling strength $J$ from observing the qubit frequency shifts in the dressed Hamiltonian. We measure the $\textit{ZZ}$ coupling strength by taking the difference of qubit B's frequency when qubit A is at $\ket{0}$ and $\ket{1}$. We fit the full model of the circuit to the data and find that $J$ can be tuned to zero at $\Phi_{\mathrm{ext}, c}/\Phi_0\sim0.3$, where $f_{01}-f_{10}$ is at its minimum. Within the range of coupler flux we measured, $J$ is tuned from $-20$MHz to $15$MHz. Over the whole range of coupler flux $\Phi_{\mathrm{ext}, c}/\Phi_{0}=0$ to $\Phi_{\mathrm{ext}, c}/\Phi_{0}=0.5$, numerical simulations show that $J$ can range between $-35$ MHz to $70$ MHz, see Fig.~\ref{fig:fig2}(a). These coupling strengths are large, of the order of individual qubit frequencies ($J \sim \omega_\mu$).

\begin{figure}[b]
\centering
\includegraphics[width=\columnwidth]{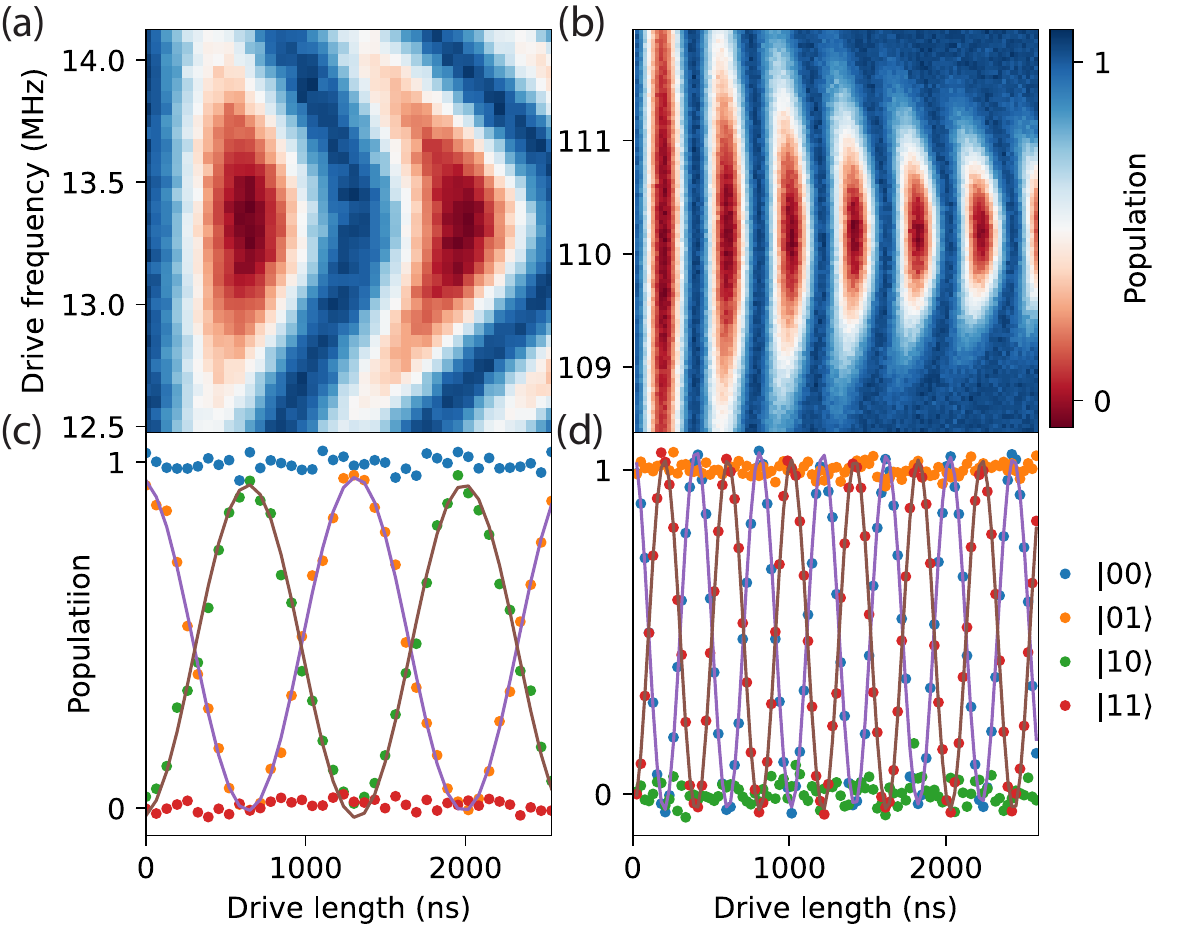}
\caption{
Sweeping coupler drive frequencies near the difference ($f_{01}-f_{10}$) and sum ($f_{01}+f_{10}$) of qubit frequencies generates chevron patterns. (a) $\ket{01}\leftrightarrow\ket{10}$ chevron centered around $13.3$MHz. (b) $\ket{00}\leftrightarrow\ket{11}$ chevron centered around $110.2$MHz. 
(c)The qubit is initialized in all 4 states, and we perform qubit A population readout after the drive. A correlated oscillation occurs between $\ket{01}$ and $\ket{10}$, while $\ket{00}$ and $\ket{11}$ populations stay constant. One full oscillation period takes $1200$ns. 
(d) correlated oscillation between $\ket{00}$ and $\ket{11}$, while $\ket{01}$ and $\ket{10}$ populations stay constant. One full oscillation period takes $400$ns.}
\label{fig:fig3}
\end{figure}

To measure the precise $\textit{ZZ}$ coupling strength, we performed Ramsey experiments on qubit B, with qubit A in its ground or excited states, see Fig ~\ref{fig:fig2}(b). We then vary the flux near the qubit sweet spots, measuring the $\textit{ZZ}$ coupling strength. As predicted by our theory, we find a $<100$Hz $\textit{ZZ}$ coupling strength point at the maximum $T_2$ point, see Fig \ref{fig:fig2}(c), implying an on-off contrast $>10^5$. At this coupler off position, qubit A (qubit B) $T_1$ is $180\mu$s ($300\mu$s), and the $T_{2e}$ is $250\mu$s ($300\mu$s), see Table~\ref{table:freqs}.

We bias the tunable coupler at the off position and realize both single and two-qubit controls with RF flux drives. Due to significant ($\sim20\%$) geometric DC and RF flux crosstalk in our system, we have off-resonant drives on both qubits. This effectively causes dynamic qubit frequency shifts as well as unwanted $\textit{ZX}, \textit{XZ}$ and $\textit{ZZ}$ drive terms (see Appendix~\ref{app:acflux}). Therefore, we develop a method to calibrate the RF flux crosstalk by measuring its effect on qubit frequencies (See Appendix~\ref{app:crosstalk}) and apply crosstalk cancellation pulses in our two-qubit gates.

To dynamically activate the $\textit{XX}$ coupling, we RF drive the coupler near the difference ($f_{01}-f_{10}$) and sum ($f_{01}+f_{10}$) frequencies of the two qubits, see Fig.~\ref{fig:fig3}. We sweep the frequency and length of this pulse at drive amplitudes of $0.5\%\Phi_0$ ($1.1\%\Phi_0$) and observe chevrons of correlated oscillations between $\ket{01}$ and $\ket{10}$ ($\ket{00}$ and $\ket{11}$), finding an oscillation rate of $0.83$MHz ($2.50$MHz). The offset from the expected frequencies is small, indicating that the crosstalk cancellation is effective.

\section{Single and two-qubit gates}
\label{sec:gates}

\begin{figure}[ht]
\centering
\includegraphics[width=\columnwidth]{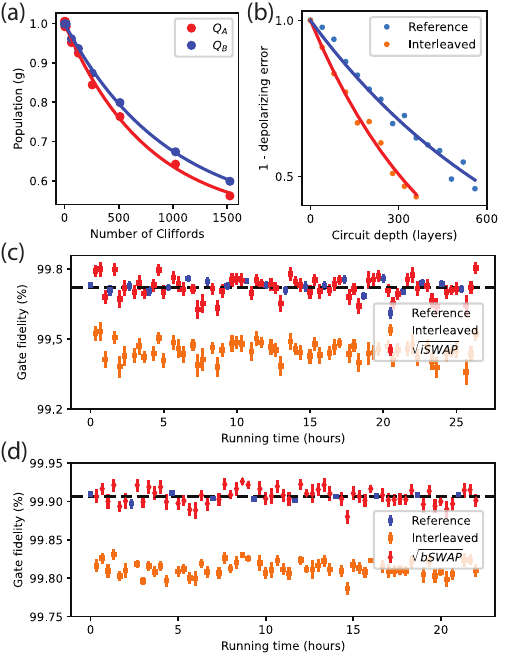}
\caption{
Measured single and two-qubit gate fidelities. (a) Simultaneous single qubit randomized benchmarking. Average single qubit Clifford gate fidelity of qubit A (qubit B) is $99.94\% (99.95\%)$. (b) Cross entropy benchmarking data, where ``Reference'' consists of layers of single-qubit gates, and ``Interleaved'' includes an interleaved $\sbswap$ gate. The depolarizing error is extracted following the procedure in~\cite{Neill2018Blueprint}. The ``Reference'' gate fidelity is $99.90\%$, and the ``Interleaved'' gate fidelity is $99.83\%$. This results in a $\sbswap$ gate fidelity of $99.93\%$. (c) $>25$ hour long consecutive cross entropy benchmarking runs for $\siswap$ gates, showing an average gate fidelity (dashed line) $99.72\%$, and error bars per point indicate statistical fit uncertainty. (d) $>20$ hour long consecutive cross entropy benchmarking runs for $\sbswap$ gates, showing an average gate fidelity (dashed line) $99.91\%$. }
\label{fig:fig4}
\end{figure}

We develop fast, high-fidelity single and two-qubit gates with our device. For the single qubit gates, we flux modulate the qubits at their corresponding bare qubit frequencies. With qubit frequencies at $48.4$ and $61.8$ MHz, we calibrated single qubit $\pi/2$ gates using Gaussian pulses with length $83.3$ and $65.1$ ns, which is approximately 4 qubit Larmor periods. We update the phases of subsequent drive pulses to implement virtual Z gates~\cite{mckay2017efficient}. We use simultaneous randomized benchmarking~\cite{Chow2009RB} to measure average single qubit Clifford gate fidelities, initially measuring qubit A (qubit B) fidelities of $99.90\% (99.88\%)$, limited primarily by coherent errors. Thus, we further optimize the single-qubit gates using DRAG shaping~\cite{Gambetta2011Analytic}, finding that the single-qubit gate fidelities increase to $99.94\%$ and $99.95\%$ respectively for each qubit, see Fig.~\ref{fig:fig4}(a). 

After pulse shaping, the remaining gate infidelity is primarily from qubit decoherence. Using a master-equation to simulate gate infidelities, we estimate contributions of qubit decay and dephasing to be $4\times10^{-4}$ and $3\times10^{-4}$, respectively. With our current pulse length, the pulse energy deviation from different carrier envelope phases is not the limiting factor of our single qubit gate fidelities. The error from such deviations is less than $1\times10^{-5}$, far lower than the dominant decoherence contribution. We also evaluated the contribution of the Bloch-Siegert shift at this drive strength to be $5.6 \times 10^{-5}$. Complete error analysis, including other sources of error, is performed in Appendix~\ref{app:error}.

For two-qubit gates, due to the negligibly small $\textit{ZZ}$ interaction strength and the tunable $\textit{XX}$ coupling, the intuitive choices for our native two-qubit gates are
\begin{align}
\siswap=\begin{pmatrix}
1 & 0 & 0 & 0 \\
0 & 1/\sqrt{2}& i/\sqrt{2} & 0 \\
0 & i/\sqrt{2} & 1/\sqrt{2} & 0 \\
0 & 0 & 0 & 1
\end{pmatrix}, \\
\sbswap=\begin{pmatrix}
1/\sqrt{2} & 0 & 0 & i/\sqrt{2} \\
0 & 1 & 0 & 0 \\
0 & 0 & 1 & 0 \\
i/\sqrt{2} & 0 & 0 & 1/\sqrt{2}
\end{pmatrix}.
\end{align}
These gates are fully entangling and have been used in demonstrating quantum advantage~\cite{arute2019quantum}. Furthermore, by embedding a $\pi$ pulse between $\siswap$ or $\sbswap$ gates, we can compile CNOT and CZ gates while also echoing out low-frequency noise. 
 
Using data shown in Fig.~\ref{fig:fig3}, we calibrate a $\siswap$ ($\sbswap$) gate that takes $257.8$ns ($101.6$ns). Because the drive frequency of $\siswap$ ($\sbswap$) is $13.4$MHz ($110.2$MHz), there are only 4 (11) oscillation periods in a single gate pulse. In the case of the $\siswap$ pulse, using a shorter gate length causes the carrier-envelope phase to affect the energy of the pulse, see Appendix~\ref{sec: error-CEP}. 

These two-qubit gates have three phase degrees of freedom, which can be calibrated by applying two virtual single-qubit $\textit{Z}_{\phi}$ gates as well as adjusting the coupler drive phase. We amplify the error associated with miscalibrated phases using specially designed sequences, allowing us to execute a pure $\siswap$ or $\sbswap$ gate. This calibration method is detailed in Appendix~\ref{app:calib}.

We measure the two-qubit gates using process tomography, but this method is limited in fidelity precision by our state preparation and measurement errors ~\cite{merkel2013self} to be $95\%$ (see Appendix \ref{app:calib}). Therefore, we perform a more precise gate fidelity measurement by amplifying gate errors and using other methods, such as cross entropy benchmarking~\cite{arute2019quantum, Boixo2018-wd} and interleaved randomized benchmarking~\cite{Magesan2012IRB}.

We implement cross entropy benchmarking by interleaving $\siswap$ ($\sbswap$) gates between random single qubit gates. To subtract the errors of the single qubit gates, we first measure a reference sequence that has the same number of layers, where each layer consists of a $\pi/2$ gate and a Z gate on each qubit, both with randomly selected phases ($\phi = n\pi/4, n \in [0, 7]$). The depolarizing error~\cite{Neill2018Blueprint, Boixo2018-wd} is found by varying the number of layers and measuring the qubit state population. Then, we repeat the process with the interleaved sequence. Because the reference gates are randomly sampled, we can calculate the depolarizing error of the two-qubit gate as $p=p_{\rm interleaved}/p_{\rm reference}$. We subsequently find the gate fidelity from the gate depolarizing error with
\begin{align}
\label{eq:depol_fid}
F=p+(1-p)/D,
\end{align}
where $p$ is the depolarizing error of the two-qubit gate, $D$ is the dimension of the Hilbert space, and $F$ is the gate fidelity. 

We continuously perform cross entropy benchmarking for both $\siswap$ and $\sbswap$ gates for more than 20 hours, as shown in Fig.~\ref{fig:fig4}. For $\siswap$, we interleaved it in reference sequences consisting of single qubit gates made of Gaussian pulses. We measured an average reference sequence depolarizing error $\sim3.2\times10^{-3}$ and an average interleaved sequence depolarizing error $\sim6\times10^{-3}$. Using equation~\ref{eq:depol_fid}, we derive an average $\siswap$ fidelity $99.72 \pm 0.04\%$. For $\sbswap$, we implemented DRAG pulses on the single-qubit gates to improve the reference sequence fidelity. We measured depolarizing error $\sim1.3\times10^{-3}$ of reference sequences and total depolarizing error $\sim2.4\times10^{-3}$ of interleaved sequences, which results in a $\sbswap$ gate fidelity $99.91 \pm 0.01\%$. 

We compute the error-budget of our single- and two-qubit gates, taking into account decoherence and other error channels, in Table~\ref{table:error_budget}. For all gates, qubit decoherence is the dominant source of error, while the beyond RWA errors and carrier envelope errors caused by using slow qubits are much lower, which shows the feasibility of high fidelity gates within the low frequency regime.

Finally we constructed a CNOT gate with two $\sbswap$ gates and five single qubit gates, and measured a CNOT fidelity of $99.5\%$ with two qubit interleaved randomized benchmarking. This result is consistent with our previously measured single qubit and $\sbswap$ gate fidelities, and can be further improved with better circuit compilation.

\section{conclusion}

We have presented a tunable coupler design for heavy-fluxonium qubits, which utilizes inductive coupling to take advantage of the large phase matrix elements. This coupler can achieve strengths that rival the single qubit energies, and can also be turned off, nulling the coupling strength to much less than the coherence time. The qubits retain high coherences and high fidelity of single qubit gates ($>99.94\%$) from simultaneous randomized benchmarking. We demonstrated fast, high fidelity $\siswap$ and $\sbswap$ gates by parametrically driving the coupler, and achieved over $99.9\%$ two-qubit gate fidelity for the $\sbswap$ gate.

In this architecture, all gate operations take place fully within the computational subspace, without occupation of higher levels. All gate operations use moderate frequency RF pulses ranging from $\sim 10-100$ MHz, that can easily be synthesized by direct digital synthesis. These demonstrated advantages can help explore a new regime of circuit design and gate schemes in the future. 

\begin{acknowledgments} 
The authors would like to thank Sho Uemera, Leandro Stefanazzi, Gustavo Cancelo for their help with the RFSoC, and Kevin He, Kan-Heng Lee, Ziqian Li, Tanay Roy, Rachel Dey for useful discussions. This work was supported by the Army Research Office under Grant No. W911NF1910016. This work is funded in part by EPiQC,
an NSF Expedition in Computing, under grant CCF1730449. This work was partially supported by the University of Chicago Materials Research Science and Engineering Center, which is funded by the National Science Foundation under award number DMR-1420709. Devices were fabricated in the Pritzker Nanofabrication Facility at the University of Chicago, which receives support from Soft and Hybrid Nanotechnology Experimental (SHyNE) Resource (NSF ECCS-1542205), a node of the National Science Foundation’s National Nanotechnology Coordinated Infrastructure. Sara Sussman was supported by the Department of Defense (DoD) through the National Defense Science \& Engineering Graduate Fellowship (NDSEG) Program.
\end{acknowledgments} 

\appendix

\section{Full circuit model}
\label{app:full_circuit}

The full Hamiltonian of the circuit shown in Fig.~\ref{fig:fig1}(b) is $H=H_{\rm qubit}+H_{\rm coupler}+V$, where
\begin{subequations}
\label{eq:fullHamiltonian}
\begin{align}
\label{eq:Hqubit}
H_{\rm qubit} &=\sum_{\mu=a,b}[4E_{C\mu}n_{\mu}^2+\frac{1}{2}E_{L\mu}\varphi_{\mu}^2+E_{J\mu}\cos(\varphi_{\mu}+\pi)] \\
\label{eq:Hcoupler}
H_{\rm coupler} &= 4E_{C-}n_{-}^2+\frac{1}{2}E_{Lc}\varphi_{-}^2-E_{Jc}\cos(\varphi_{-}+2\pi\Phi_{c}/\Phi_{0}) \\ \nonumber 
&\quad+ 4E_{C+}n_{+}^2+\frac{1}{2}E_{Lc}\varphi_{+}^2, \\
\label{eq:V}
V&=-\frac{E_{La}}{2}[\varphi_{a}(\varphi_{+}+ \varphi_{-})]
-\frac{E_{Lb}}{2}[\varphi_b(\varphi_{+}- \varphi_{-})] \\ \nonumber 
&\quad+\sum_{\mu=a,b}\frac{E_{L\mu}}{2}{\delta\phi_{\mu}}[-2\varphi_{\mu}+\varphi_{+}+(-1)^{\mu}\varphi_{-}],
\end{align}
\end{subequations}
where $[\varphi_{\mu}, n_{\nu}]=i\delta_{\mu\nu},\mu=a,b,-,+$, and we have defined $E_{Lc}=\frac{1}{2}(\frac{1}{2}[E_{La}+E_{Lb}]+E_{L})$, $E_{C+}=2E_{Cc},$ and $E_{C-}=(1/E_{Cm}+1/[2E_{Cc}])^{-1}$. All other circuit parameters can be read off from Fig.~\ref{fig:fig1}(b).
To account for the need to tune the qubit flux as a function of the coupler flux as discussed in the main text, we define $\delta\phi_{\mu}=\phi_{\mu}-\pi$.
See Ref.~\cite{Weiss2022} for further details on the derivation of Eq.~\eqref{eq:fullHamiltonian}.

We comment briefly on how the effective Hamiltonian Eq.~\eqref{eq:effHam} arises from Eq.~\eqref{eq:fullHamiltonian}. The $\textit{XX}$ operator content of the effective coupling between the qubits can be traced back to the operator form of the coupling between the qubits and the couplers $\varphi_{\mu}\varphi_{\pm}$. The phase operator in the qubit subspace of a fluxonium qubit biased at the sweet spot can be written as $\langle 1|\varphi|0\rangle\sigma_{x}$ \cite{Zhang2021Universal}. Thus, one application of the perturbing term $\varphi_{a}\varphi_{\pm}$ swaps e.g. an excitation of qubit $a$ into the coupler, and an application of $\varphi_{b}\varphi_{\pm}$ returns this excitation to the qubit subspace in the form of an excitation of qubit $b$. For further details, we refer the reader to Ref.~\cite{Weiss2022}.

Using this Hamiltonian we locate the sweet-spot contour by minimizing the eigenenergy $E_{1100}$ as a function of the qubit fluxes, keeping the coupler-flux fixed. The off position is then found by performing the same minimization along the sweet-spot contour as a function of coupler flux.

\section{Plasmon spectroscopy and full model fitting}
\label{app:spectroscopyfit}

\begin{figure}[h]
\centering
\includegraphics[width=\columnwidth, trim=0 0 0.5in 0.5in, clip]{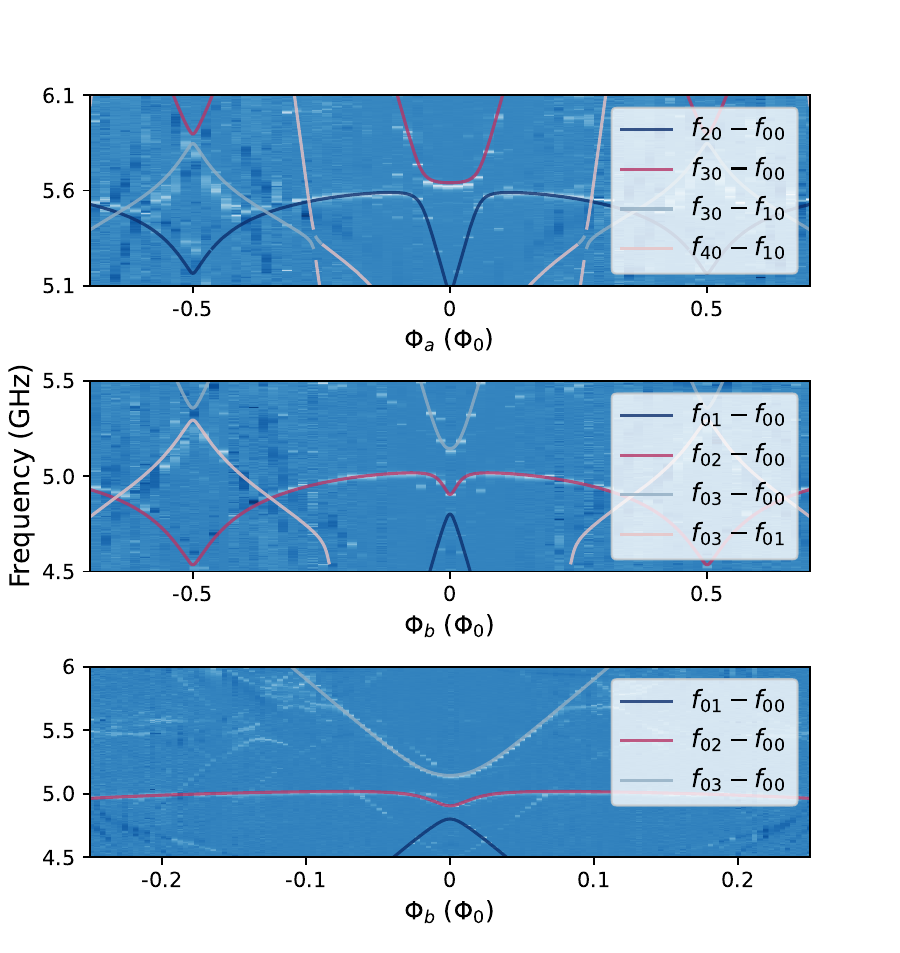}
\caption{
Plasmon spectroscopy features of both qubits. Top panel: $\ket{00}\rightarrow\ket{20}$ and $\ket{00}\rightarrow\ket{30}$ transition frequency vs qubit A flux bias. Middle panel: $\ket{00}\rightarrow\ket{02}$ and $\ket{00}\rightarrow\ket{03}$ transition frequency vs qubit B flux bias. Bottom panel: zoomed in plot around $\Phi_{\mathrm{ext}, b}=0$. The data was taken with coupler flux around 0, and the other qubit's flux fixed at 0.}
\label{fig:plasmon_spec}
\end{figure}

To determine our system's physical parameters, we measure the two-tone spectroscopies of higher ($\sim5$GHz) frequency transitions. Each qubit is charge-driven through the readout resonator while the resonator is being probed. We see a sharp change in the resonator transmission when a qubit transition is driven on resonance. Thus, by sweeping different flux biases, we can see a frequency-flux 2D qubit spectroscopy. We measured the transition frequencies from the ground state to first and second plasmon states ($\ket{20}$ and $\ket{30}$) for qubit A while changing the qubit A flux bias. Because $\ket{10}$ has significant thermal population around the $\Phi_{\mathrm{ext},a}=0.5\Phi_0$ bias point, we could also see transitions from it to the higher levels (see Fig.~\ref{fig:plasmon_spec}(Top). Similarly, we measured transition frequencies for qubit B (see Fig.~\ref{fig:plasmon_spec}(Middle and Bottom). Using the package \texttt{scqubits}~\cite{groszkowski2021scqubits, chitta2022computer}, we subsequently fitted the spectroscopy data with the full Hamiltonian Eq.~\eqref{eq:fullHamiltonian}, as shown in overlay lines in Fig.~\ref{fig:plasmon_spec}. Because our DC flux crosstalk matrix is not perfectly calibrated across this range, we could not keep the coupler and the other qubit flux biases precisely at the same point during a flux sweep, leading to slight mismatches between the theory curves and experimental data.

\begin{table}[]
\caption{\label{table:freqs} Parameters and basic properties for the qubits. Measurements are made when the fluxonium qubits are biased at their half-flux sweet spots while the coupler $\varphi_{-}$ mode is biased at $\Phi_{\mathrm{ext},c}/\Phi_{0}=0.3$. The energy splittings are differences between the bare ground and first-excited state energies $f_{10}=f_{1}-f_{0}$, while the anharmonicity is defined as $\alpha=f_{21}-f_{10}$. Circuit parameters in GHz used throughout this work. The measured $T_2$ in this table are the values when the other qubit is at its ground state.}

\begin{tabularx}{\linewidth}{*9{>{\centering\arraybackslash}X}}
             & qubit $a$ & qubit $b$ & coupler $\varphi_{-}$ & coupler $\varphi_{+}$ \\ \hline \hline
$f_{10}$ (GHz) & 0.0618    & 0.0484    & 9.52                   & 15.6                   \\
$\alpha$ (GHz) & 4.41       & 5.06       & 0.93                 & 0                    \\ \hline \hline
$T_1$ ($\mu$s) & 180 & 300 \\
$T_2^*$ ($\mu$s) & 150 & 200 \\
$T_{2e}$ ($\mu$s) & 250 & 300 \\ \hline \hline 
$E_J$ (GHz)    & 5.65      & 4.88      & 4.246                &                      \\
$E_C$ (GHz)    & 0.95      & 0.905     & 8                    & 12                   \\
$E_L$ (GHz)    & 0.292     & 0.286     & 3.52 & 3.52                    \\
               &           &           &                      &                     
\end{tabularx}
\end{table}

\section{\textit{ZZ} suppression}
\label{app:ZZ}

The suppression of static \textit{ZZ} coupling is an attractive feature of our architecture, allowing for large on/off ratios. This suppression is achieved by cancelling out two contributions of $ZZ$ by fine tuning the qubit flux bias. To see this, we start from the effective Hamiltonian 
\begin{align}
\label{eq:effHam_appZZ}
H_{\text{eff}} &= -\sum_{\mu=a,b}\frac{\omega_{\mu}}{2}\sigma_{z}^{\mu}-\Omega_{\mu}\sigma_{x}^{\mu}+ J\sigma_{x}^{a}\sigma_{x}^{b}+\zeta\sigma_{z}^{a}\sigma_{z}^{b},
\end{align}
where Pauli operators are defined in the eigenbasis of the system at the off position. The static $ZZ$ shift $\zeta$ is small due to a number of reasons. First, $\zeta$ is generally suppressed in low-frequency fluxonium architectures, due to the large detuning between the qubit excitation energies and the energies of the higher-lying states ~\cite{ficheux2021fast}. Second, the galvanic-coupling architecture helps suppress $\zeta$ because the phase matrix elements of qubit states with higher-lying fluxonium states are very small relative to charge matrix elements. Finally, there is no direct coupling between the qubits in this architecture . Thus, second- and third-order contributions to $\zeta$ vanish, with the leading-order contribution arising only at fourth order in perturbation theory~\cite{Sung2021, zhu2013circuit, ding2023high}. Sweeping the coupler flux while keeping qubits at their sweet spots ($\Omega_\mu = 0$), $|\zeta|$ is nonvanishing and on the order of 1 kHz. 

\begin{figure}[h]
\centering
\includegraphics[width=\columnwidth]{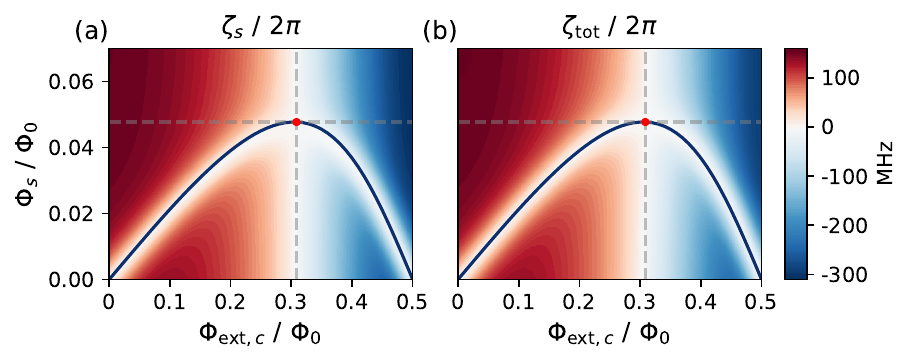}
\caption{
The $ZZ$ interaction strength (a) $\zeta_{s}$ obtained from the effective model and (b) $\zeta_{\mathrm{tot}}$ from numerical diagonalization of the full model. Both quantities are plotted as a function of the qubit flux shifts $\Phi_s\equiv\Phi_{\mathrm{ext}, a} - 0.5\Phi_0 = 0.5\Phi_0 - \Phi_{\mathrm{ext}, b}$ and the coupler flux $\Phi_{\mathrm{ext}, c}$. In both plots, the red circle marks the off position and the solid curves are the ``sweet-spot contours''. We find that $\zeta_s$ is the main contributor to $\zeta_{\mathrm{tot}}$, and $\zeta_s$ can be tuned from positive to negative values to cancel out $\zeta$. 
}
\label{fig:zzscan}
\end{figure}

However, $\zeta$ can be canceled with a non-zero $\Omega_\mu$ by qubit-flux dependent shifts away from the sweet-spot contour. 
To see this, we diagonalize the single-qubit terms in Eq.~\eqref{eq:effHam_appZZ} up to second order of $\Omega_\mu/\omega_\mu$ using the unitary transformation $U=\exp(-i\sum_{\mu=a,b}\frac{\Omega_{\mu}}{\omega_{\mu}}\sigma_{y\mu})$, yielding
\begin{align}
\label{eq:Heff_appZZ_p}
H_{\mathrm{eff}}' &= U^{\dagger}H_{\rm eff}U \\ \nonumber 
 &=-\sum_{\mu=a,b}\frac{\omega_{\mu}+2\frac{\Omega_{\mu}^2}{\omega_{\mu}}}{2}\sigma_{z}^{\mu} \\ \nonumber 
&\quad+ J\left(\sigma_{x}^{a}+\frac{2\Omega_{a}}{\omega_{a}}\sigma_{z}^{a}\right)\left(\sigma_{x}^{b}+\frac{2\Omega_{b}}{\omega_{b}}\sigma_{z}^{b}\right) + \zeta \sigma_{z}^{a}\sigma_{z}^{b},
\end{align}
where we have expanded the result to second order in the small parameters $\Omega_{\mu}/\omega_{\mu}$ and ignored negligible corrections arising from the last term in Eq.~\eqref{eq:Heff_appZZ_p}. Thus the overall $\textit{ZZ}$ shift is given by 
\begin{align}
\label{eq:zetaprime}
\zeta_{\mathrm{tot}}=\zeta + \zeta_s, \quad  \mathrm{where }\ \zeta_{s} = J \frac{4\Omega_{a}\Omega_{b}}{\omega_{a}\omega_{b}}
\end{align}
where both contributions effectively arise as fourth order perturbations. The sign of $\zeta_{s}$ depends upon the signs of $J$ and $\Omega_{\mu}$, where notably $J$ changes sign when moving across the off position in coupler flux, see Fig.~\ref{fig:zzscan}(a). 
Thus with small adjustments in the qubit fluxes (so that the $\Omega_{\mu}$ are nonzero) and the coupler flux, $\zeta_s$ can be used to cancel $\zeta$. This leads to an overall vanishing $\textit{ZZ}$ shift $\zeta_{\mathrm{tot}}=0$.

We validate this understanding by calculating $\zeta_{\mathrm{tot}}$ using the full Hamiltonian \eqref{eq:fullHamiltonian}
\begin{align}
\label{eq:ZZ_4}
\zeta_{\mathrm{tot}} / 2\pi = f_{11}-f_{01}-f_{10}+f_{00},
\end{align}
see Fig.~\ref{fig:zzscan}(b). The off position and $\zeta_{\mathrm{tot}}=0$ point are within $10^{-5} \Phi_{0}$ in the qubit fluxes according to the full model, relaxing the constraint of operating exactly on the sweet-spot contour. We need not worry about tuning of the qubit fluxes away from the sweet spot, as flux tuning at the $10^{-5} \Phi_{0}$ level is below the resolution of our DC bias source and contributes minimally to flux-noise induced dephasing. 

\section{Effects due to RF flux crosstalk}
\label{app:acflux}
In the presence of RF flux crosstalk where flux penetrates each of the qubit loops when we drive the coupler, dynamic effects arises that reduce the gate fidelity. For simplicity, we ignore the effects of the pulse's Gaussian envelope in this section. The Hamiltonian of the system is

\begin{align}
\label{eq:H_cross}
H_{\rm eff,ct} =& - \sum_{\mu=a,b} \left( \frac{\omega_\mu}{2}\sigma_{z}^\mu + \Omega_{\mathrm{ct}, \mu} \cos(\omega_{d}t)\sigma_{x}^\mu  \right) \\ \nonumber 
&+ J_{\mathrm{ac}}\cos(\omega_{d}t)\sigma_{x}^{a}\sigma_{x}^{b} 
\end{align}
where $\Omega_{\mathrm{ct}, \mu}$ describes the maximum qubit-flux drive amplitude due to crosstalk. The drive frequency will typically be set to $f_{d}\approx f_{01}-f_{10}$ for an $\siswap$-style gate, or $f_d \approx f_{01}+ f_{10}$ for a $\sbswap$-style gate, with $\omega_d = 2\pi f_d$. In each case, the drives are detuned from single-qubit transition frequencies. This unwanted drive thus dynamically shifts the qubit frequencies rather than inducing Rabi oscillations~\cite{Blais2007}. To see this, we first move to a frame rotating at the drive frequency. Making the RWA for the single-qubit drive terms while keeping all terms associated with the two-qubit drive, we obtain
\begin{align}
\label{eq:ac H'}
H_{\rm eff,ct}' &= -\sum_{\mu=a,b} \left(\frac{\omega_{\mu}-\omega_{d}}{2}\sigma_{z}^{\mu} + \frac{\Omega_{\mathrm{ct}, \mu}}{2}\sigma_{x}^\mu \right) \\ \nonumber 
&\quad+J_{\mathrm{ac}}\cos(\omega_{d}t)(\cos(\omega_{d}t)\sigma_{x}^{a}+\sin(\omega_{d}t)\sigma_{y}^{a})
\\ \nonumber &\qquad\times(\cos(\omega_{d}t)\sigma_{x}^{b}+\sin(\omega_{d}t)\sigma_{y}^{b}),
\end{align}
As in Sec.~\ref{app:ZZ}, we diagonalize the single-qubit terms in Eq.~\ref{eq:ac H'} to second-order in $\Omega_{\mathrm{ct}, \mu} / (\omega_{\mu}-\omega_{d})$, yielding
\begin{align}
&H_{\rm eff,ct}'' = -\sum_{\mu=a,b}\frac{\omega_{\mu} + \delta\omega_{\mu}}{2}\sigma_{z}^{\mu}  \\ \nonumber 
&+J_{\mathrm{ac}}\cos(\omega_{d}t)\left( \sigma_{x}^{a} + \frac{\Omega_{\mathrm{ct}, a}}{\omega_{a}-\omega_{d}}\sigma_{z}^{a} \right) 
 \left( \sigma_{x}^{b} + \frac{\Omega_{\mathrm{ct}, b}}{\omega_{b}-\omega_{d}}\sigma_{z}^{b} \right),
\end{align}
where $\delta\omega_{\mu} = \Omega_{\mathrm{ct}, \mu}^2 / 2(\omega_{\mu}-\omega_{d})$ and we have returned to the lab frame.
Thus, the leading-order effects of ac-flux crosstalk are to renormalize the qubit frequencies as well as modify the effective two-qubit drive term from $XX$ to a combination of $XX, ZX, XZ$ and $ZZ$. To avoid these unwanted dynamic effects, it is thus critical to cancel geometric flux crosstalk. We discuss the quantitative contribution of ac-flux crosstalk to gate infidelities in Appendix~\ref{app:error}.

\section{Experimental setup}
\label{app:wiring}
\begin{figure*}[t]
\centering
\includegraphics[width=\textwidth]{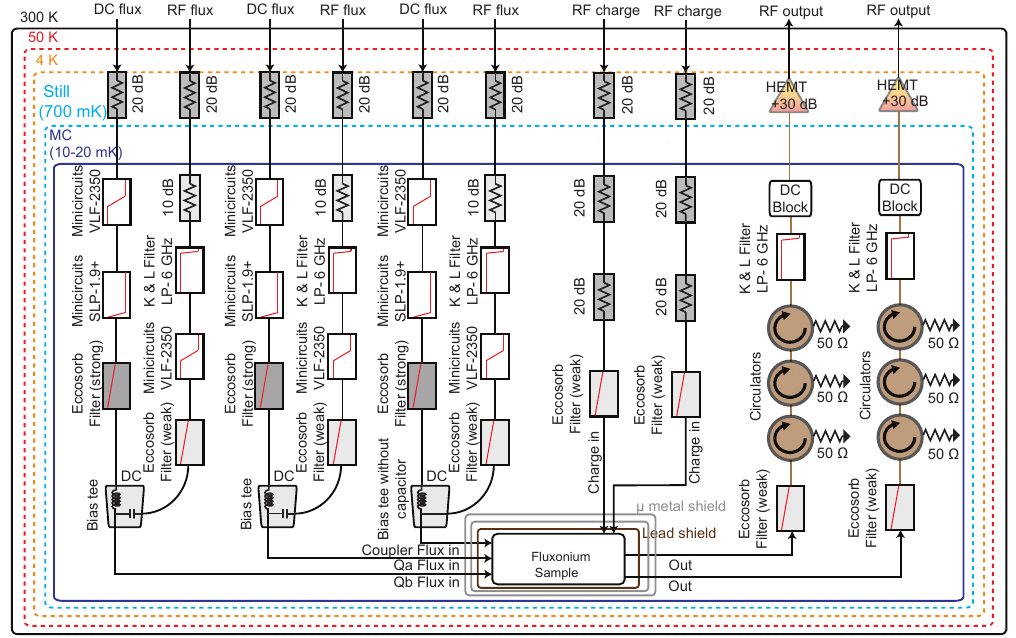}
\caption{Wiring diagram inside the dilution refrigerator. Outside the dilution fridge, there is $13-20~$ dB of attenuation and a DC block on the RF flux line, and an ultra low pass ($< 1~$ Hz) RC filter on the DC flux line. Adding attenuation on the RF flux lines at room temperature led to a measured increase in the $T_1$ and $T_2$ coherences of the qubits.}
\label{fig:setup}
\end{figure*}
The experiment was performed in a Bluefors LD-250 dilution refrigerator with the wiring configured as shown in Fig.~\ref{fig:setup}. The flux and charge inputs are attenuated at the $4$K stage and the mixing chamber with standard XMA attenuators, except the final $20$dB attenuator on the RF charge line (threaded copper). The DC and RF-flux signals were combined in a bias-tee (Mini-Circuits\textsuperscript{\textregistered} ZFBT-4R2GW+), where the coupler bias-tee was modified such that the capacitor is replaced with a short to further lower the high-pass cutoff frequency.  The DC and RF-flux lines included commercial low-pass filters (Mini-Circuits\textsuperscript{\textregistered}) as indicated. The RF flux and output lines also had additional low-pass filters with a sharp cutoff (8 GHz) from K\&L microwave. Home-made Eccosorb (CR110) IR filters were added on the flux, input and output lines, which further improved the $T_1$ and $T_2$ coherences, and reduced the qubit and resonator temperatures. The device was heat sunk to the base stage of the dilution refrigerator (stabilized at 15~mK) via an OFHC copper post, while surrounded by an inner lead shield thermalized via a welded copper ring. This was additionally surrounded by two cylindrical $\mu$-metal cans (MuShield), thermally anchored using an inner close fit copper shim sheet, attached to the copper can lid. We ensured that the sample shield was light tight, to reduce thermal photons from the environment.

\section{Device fabrication}
\label{app:fab}
The device (shown in Fig.~\ref{fig:fig1} in the main text) was fabricated on a 430$\mu$m thick C-plane sapphire substrate. The base layer of the device, which includes the majority of the circuit (excluding the Josephson junctions), consists of 200nm of tantalum with features fabricated via optical lithography and HF etch at wafer-scale. 

We perform standard TAMI cleaning for annealed sapphire substrates, followed with nanostrip etching at 50C for 10 minutes, and sulfuric acid etching at 140C for 10 minutes. We subsequently deposit 200nm tantalum in an AJA ATC 2200 sputtering tool at 800C. A 2000nm thick layer of AZ 1518 was used as the (positive) photoresist, and the large features were written using a Heidelberg MLA 150 Direct Writer. After 60 seconds of development with MIF 300 and a 10 minute oven bake at 120C, we perform 20 seconds of HF etching using Ta etchant 1:1:1 (Transene Tantalum Etchant 111). 

The junction mask was fabricated via electron-beam lithography with a bi-layer resist (MMA-PMMA) comprising of MMA EL11 and 495PMMA A6 spun at 4000RPM. The e-beam lithography was performed on a 100kV Raith EBPG5000 Plus E-Beam Writer. All Josephson junctions were made with the Dolan bridge technique and etched for 2 minutes using a 3:1 DI:IPA solution at 6C. Aluminum was subsequently evaporated onto the chip in a  Plassys electron beam evaporator using double angle evaporation ($\pm 21^o$), first depositing a 70nm Al layer, performing static oxidation, and subsequently depositing a 90nm Al layer. The wafer was then diced into $7\times7$ mm chips, mounted on a printed circuit board, and subsequently wire-bonded.

\section{Readout and active reset}
\label{app:readout}

\begin{figure}[h]
\centering
\includegraphics[width=\columnwidth]{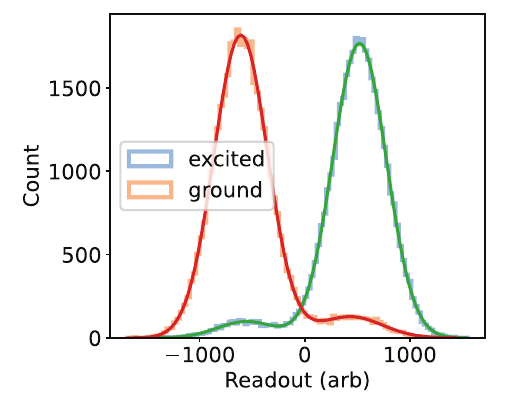}
\caption{
Histogram of readout data after initializing the qubit in either ground or excited state via active reset. The populations are then fit to a double-Gaussian function and subsequently plotted as solid red and green solid lines. This fit is analyzed to determine the qubit temperature. }
\label{fig:histogram}
\end{figure}

We utilize the active feedback capabilities of the Xilinx RFSoC FPGA board with the QICK open source control codebase. We perform simultaneous dispersive readout, using readout lengths of $10.4$ and $18.2\mu$s respectively. The readout data will then be processed on the FPGA board and compared to a previously calibrated threshold, and the board will conditionally play a single qubit $\pi$ pulse to initialize the qubit to its ground state. 

We first take $10000$ readouts with the qubit uninitialized at the sweet spot. Because the qubit's Boltzmann temperature ($<2.8$mK) is lower than the temperature of the environment, we expect to see the qubit in thermal equilibrium. Indeed, we see approximately a 50-50 population split between the ground and excited states. Using a double Gaussian function, we fit the data to determine the demarcation line that best classifies the qubit state for use in readout and active reset. 

We measure our reset fidelity by initializing the qubit in $\ket{g}$ and $\ket{e}$ using active reset, and then take another measurement right after the reset to get histogram shown in Fig.~\ref{fig:histogram}. We extracted post-selected active reset fidelities of $93\%$ and $95\%$ respectively for each qubit, and it is very close to the readout fidelities due to the fast feedback time and high fidelity single qubit $\pi$ pulse. By fitting the thermal population of the qubit ground and excited states, we can also measure the qubit temperatures to be $\sim50$mK.

\section{Flux crosstalk calibrations}
\label{app:crosstalk} 

\begin{figure}[h]
\centering
\includegraphics[width=\columnwidth]{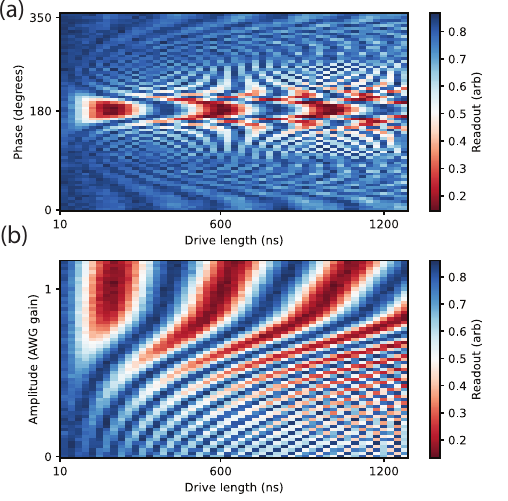}
\caption{
(a) Performing a drive at the sum of qubit frequencies with various lengths while sweeping the phase of a cancellation pulse. The amplitude of this sweep is randomly chosen, but it needs to be lower than the correct cancellation amplitude. (b) with the phase calibrated in (a), performing a drive at the sum of qubit frequencies with various lengths while sweeping the amplitude of a cancellation pulse.}
\label{fig:flux_crt}
\end{figure}

The DC and RF flux crosstalk of the system are significant and require calibration to mitigate unwanted effects that degrade gate fidelity. We measure the DC flux crosstalk by performing 2D flux sweeps for different pairs of flux lines while probing the resonator. This reveals sharp flux-dependent features that originate from bringing qubit energy levels on resonance with the readout resonator. The slope of these lines is the DC flux crosstalk of our system.

To minimize RF flux crosstalk, we use compensation pulses that are played at the same time as the original drive pulse. These compensation pulses have the same length as the original pulse, but we tune the phase and amplitude of these pulses to minimize the effect of flux crosstalk. Since the phases of these compensation pulses are independent parameters, we first calibrate them individually. Because the flux crosstalk can be understood as an off-resonant qubit drive, we would observe an AC Stark-shifted qubit frequency when performing a Rabi experiment. We Rabi drive at the $\sbswap$ frequency, which is the sum of bare qubit frequencies. The Rabi contrast is maximized when the qubit frequency shift is minimized, making this a good metric for measuring RF crosstalk. Therefore, we sweep the phase of the compensation pulse while playing it simultaneously with the drive pulse, as shown in Fig~\ref{fig:flux_crt}(a). The maximum Rabi contrast of the $\ket{00}\leftrightarrow\ket{11}$ oscillation determines the correct compensation pulse phase.

After fixing the correct phases for both compensation channels, we determine the optimal cancellation pulse amplitudes in a similar sweep. As before, we look for the maximum Rabi contrast of the $\ket{00}\leftrightarrow\ket{11}$ oscillation, but in this case, we sweep the amplitude of the compensation pulse, shown in Fig~\ref{fig:flux_crt}(b). Because the cancellation gains from the two qubit channels will affect each other, we did this iteratively to find the optimal amplitudes for both of them. 

Due to the linear nature of the flux crosstalk, the amplitude of drive pulse and compensation pulses have a linear relationship, and the phase difference between all three pulses is not amplitude dependent. With cancellation phases and amplitudes for one drive pulse calibrated, we can utilize this procedure to find compensation pulses for all $\ket{00}\leftrightarrow\ket{11}$ drive pulses. The two-qubit gates reported in this manuscript are all constructed in this way, with the coupler drive pulse and cancellation pulses on each qubit.

\section{two-qubit gate calibrations}
\label{app:calib}

\begin{figure}[h]
\centering
\includegraphics[width=\columnwidth]{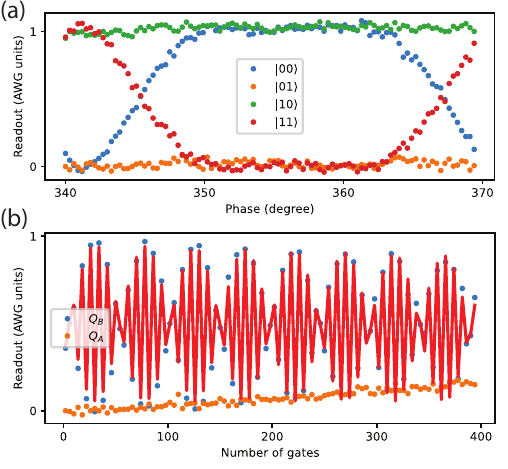}
\caption{
(a) Playing the pulse sequence in Fig.~\ref{fig:calibcirq1} while sweeping the phase of qubit A phase gates ($Z_A$). At $\phi_A=357\degree$, the phase of $\phi_{01}+\phi_{10}$ is perfectly cancelled, thus $\phi_{01}+\phi_{10}=3$. (b) Playing the pulse sequence in Fig.~\ref{fig:calibcirq2} with various copies of the $\sbswap$ gates, and measure the final state of both qubits. With this sequence, the qubit A should only see a decay process to thermal equilibrium, and qubit B state dependence on the number of gates is shown in Eqs.~\ref{eq:phasecalib}. By fitting the data we can extract the value of $\phi_{01}$ and $\phi_{10}$.}
\label{fig:phase_calib}
\end{figure}

We calibrate the $\siswap$ and $\sbswap$ gates with gate sequences designed to amplify the gate errors. The gate calibration process consists of two steps, the rotation angle calibration and phase calibrations.
We can write the Hermitian matrix for a gate generated by a generic $\textit{XX}$ parametric drive on the coupler as
\begin{align}
\sqrt{\phi_b\mathrm{SWAP}}=
\begin{pmatrix}
\cos{\theta} & 0 & 0 & i e^{i\phi_D}\sin{\theta} \\
0 & e^{i\phi_{01}} & 0 & 0 \\
0 & 0 & e^{i\phi_{10}} & 0 \\
i e^{i(\phi_{11}-\phi_D)}\sin{\theta} & 0 & 0 & e^{i\phi_{11}}\cos{\theta}
\end{pmatrix}
\end{align}

for $\ket{00}\leftrightarrow\ket{11}$ oscillation and

\begin{align}
\sqrt{\phi_i\mathrm{SWAP}}=
\begin{pmatrix}
1 & 0 & 0 & 0 \\
0 & e^{i\phi_{01}}\cos{\theta} & i e^{i(\phi_{01}+\phi_D)}\sin{\theta} & 0 \\
0 & i e^{i(\phi_{10}-\phi_D)}\sin{\theta} & e^{i\phi_{10}}\cos{\theta} & 0 \\
0 & 0 & 0 & e^{i\phi_{11}}
\end{pmatrix}
\end{align}

for $\ket{01}\leftrightarrow\ket{10}$ oscillation, where $\phi_D$ is the phase of the coupler drive, $\phi_{01}$, $\phi_{10}$, and $\phi_{11}$ are phases due to the frequency shift of levels while the drive is on, which have relationship $\phi_{11}=\phi_{01}+\phi_{10}+\phi_{zz}$~\cite{ganzhorn2020benchmarking}. In our system, since all frequency shifts and the $\textit{ZZ}$ term during the parametric drive are small, we have $\phi_{11}\approx\phi_{01}+\phi_{10}$ and all of $\phi_{01}$, $\phi_{10}$, and $\phi_{11}$ are close to 0. 
We first calibrate the rotation angle ($\theta$). We fixed the pulse length for $\siswap$ at $99000/384\approx257.8$ns and $\sbswap$ at $39000/384\approx101.6$ns. Since our AWG (Xilinx RFSoC) has a $384$MHz processor, the pulse lengths are integer multiples of $1000/384$ns. We initialize the system in state $\ket{00}$, sweep the pulse amplitude while playing the same pulse consecutively for $4n+2$ times, and look for the amplitude that gives us the highest fidelity $\ket{11}$ state. With playing the same gate for up to 402 times, we can obtain the value of parameter $\theta$ up to $1\times10^{-4}$ in precision.

\begin{figure}[h]
\centering
\includegraphics[width=\columnwidth]{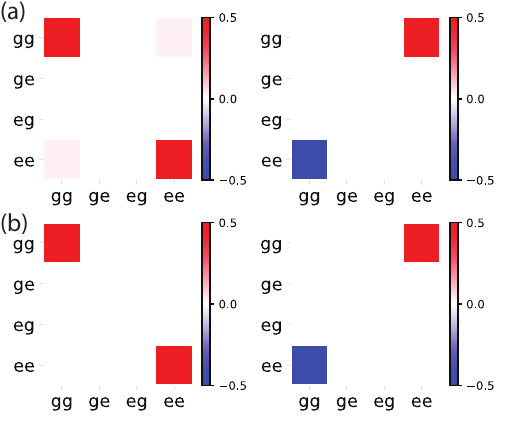}
\caption{
(a) Experimental data of state tomography on $\ket{gg} - i\ket{ee}$. This state was prepared by directly playing a $\siswap$ gate on an initialized $\ket{gg}$ state. On the left is the real part, and on the right is on the imaginary part.(b) State tomography of an ideal $\ket{gg} - i\ket{ee}$ calculated with theory. }
\label{fig:statetomo}
\end{figure}

With $\theta$ calibrated to be $\pi/4$, gate phases $\phi_{01},\phi_{10},\phi_{11}$ and $\phi_D$ are subsequently calibrated. Since the calibration process is very similar for $\siswap$ and $\sbswap$ gates, here we only use $\sbswap$ as an example for explanation. We calibrate these phases with the sequence shown in Fig.\ref{fig:calibcirq1}: 

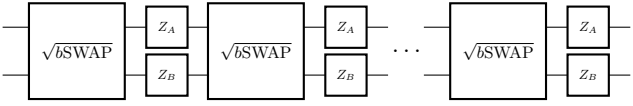
\begin{figure}[h]\centering
\begin{circuitikz}[scale=0.7]
\draw[scale=1.0,transform shape] (0,0)node[fourport,t={$\sbswap$}](tg1){};
\draw[scale=0.8,transform shape] (tg1.2) to[twoport,t={$Z_B$}] ++(2,0);
\draw[scale=0.8,transform shape] (tg1.3) to[twoport,t={$Z_A$}] ++(2,0);
\draw(tg1.1) to[short] ++(-0.5,0)
(tg1.4) to[short] ++(-0.5,0);
\draw[scale=1.0,transform shape] (3.4,0)node[fourport,t={$\sbswap$}](tg2){};
\draw[scale=0.8,transform shape] (tg2.2) to[twoport,t={$Z_B$}] ++(2,0);
\draw[scale=0.8,transform shape] (tg2.3) to[twoport,t={$Z_A$}] ++(2,0);
\path (5.8,0) -- node[auto=false]{\ldots} (6.8,0);
\draw[scale=1.0,transform shape] (8,0)node[fourport,t={$\sbswap$}](tg3){};
\draw(tg3.1) to[short] ++(-0.5,0)
(tg3.4) to[short] ++(-0.5,0);
\draw[scale=0.8,transform shape] (tg3.2) to[twoport,t={$Z_B$}] ++(2,0);
\draw[scale=0.8,transform shape] (tg3.3) to[twoport,t={$Z_A$}] ++(2,0);
\end{circuitikz}
\caption{Gate sequence for calibrating $\sbswap$ gate phase $\phi_{11}$.}
\label{fig:calibcirq1}
\end{figure}

where $Z_A$ and $Z_B$ are single qubit phase gates on each qubit with phase $\phi_A$ and $\phi_B$. A unit consists of one $\sbswap$ gate and two single qubit Z gates, and we noticed that with $4n+2$ number of units, only when $\phi_A+\phi_B=-\phi_{11}$ do we get a bSWAP gate up to some phases. Thus we play a sequence of 321 units while sweeping $\phi_A$ (shown in fig~\ref{fig:phase_calib}), and measured $\phi_{11}$ to be around $-3$ degrees. We subsequently play the sequence as shown in Fig.\ref{fig:calibcirq2}: 

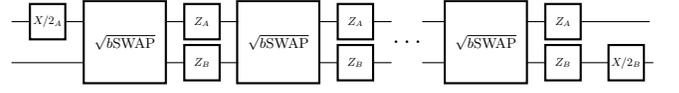
\begin{figure}[h]\centering
\begin{circuitikz}[scale=0.6]
\draw[scale=1.0,transform shape] (0,0)node[fourport,t={$\sbswap$}](tg1){};
\draw[scale=0.8,transform shape] (tg1.2) to[twoport,t={$Z_B$}] ++(2,0);
\draw[scale=0.8,transform shape] (tg1.3) to[twoport,t={$Z_A$}] ++(2,0);
\draw[scale=0.8,transform shape] (tg1.4) to[twoport,t={$X/2_A$}] ++(-2,0);
\draw (tg1.1) to[short] ++(-1.6,0);
\draw[scale=1.0,transform shape] (3.4,0)node[fourport,t={$\sbswap$}](tg2){};
\draw[scale=0.8,transform shape] (tg2.2) to[twoport,t={$Z_B$}] ++(2,0);
\draw[scale=0.8,transform shape] (tg2.3) to[twoport,t={$Z_A$}] ++(2,0);
\path (5.8,0) -- node[auto=false]{\ldots} (6.8,0);
\draw[scale=1.0,transform shape] (8,0)node[fourport,t={$\sbswap$}](tg3){};
\draw(tg3.1) to[short] ++(-0.5,0)
(tg3.4) to[short] ++(-0.5,0);
\draw[scale=0.8,transform shape] (tg3.2) to[twoport,t={$Z_B$}] ++(2,0)
 to [twoport,t={$X/2_B$}] ++(1.5,0);
\draw[scale=0.8,transform shape] (tg3.3) to[twoport,t={$Z_A$}] ++(2,0)
 to [short] ++(1.4,0);
\end{circuitikz}
\caption{Gate sequence for calibrating $\sbswap$ gate phases $\phi_{10}$ and $\phi_{D}$.}
\label{fig:calibcirq2}
\end{figure}

and with $4n+2$ units, we have qubit B ground state population,

\begin{align}
\frac{1}{2}(1+(-1)^n\sin{(4n(\phi_{10}-\phi_{11})+2(\phi_{10}-\phi_{11})+\phi_D}))
\label{eq:phasecalib}
\end{align}
and qubit A ground state population should always be 0 for all the n values. Fitting this function gives us $\phi_{10}-\phi_{11}$ and $2\phi_{10}-2\phi_{11}+\phi_D$ (shown in figure~\ref{fig:phase_calib}). With the $\phi_{11}$ value measured from the step above, we can derive $\phi_{10}$, $\phi_{11}$, and $\phi_D$. Since $\phi_{ZZ}$ is very small in our system, we can calculate $\phi_{01}$ as $\phi_{11}-\phi_{10}$, and the result is very close to directly measuring $\phi_{01}$ by switching operations on qubit A and qubit B.

After the calibration, we set qubit A Z gate after the parametric coupler drive $\phi_A=-\phi_{10}$, qubit B Z gate $\phi_B=-\phi_{01}$, and change the drive phase by $-\phi_D$. In this experiment all the single qubit Z gates are virtual, therefore we only need to update the phase of the pulses after a $\sbswap$ or $\siswap$ gate for each qubit.

\begin{figure}[h]
\centering
\includegraphics[width=\columnwidth]{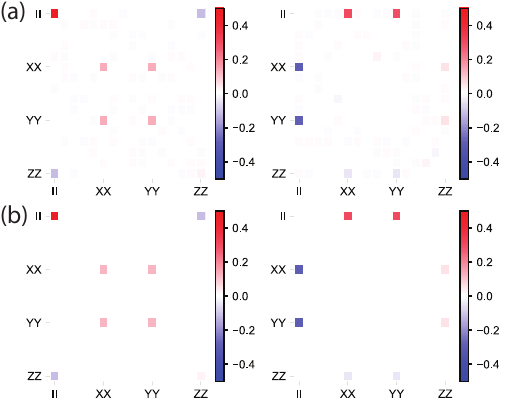}
\caption{
(a) Experimental data of process tomography on $\siswap$ gate, where the left side shows the real part and the right side shows the imaginary part. (b) Process tomography of an ideal $\siswap$ gate calculated with theory.}
\label{fig:processtomo}
\end{figure}

\section{Error budgeting}
\label{app:error}

Various sources contribute to single- and two-qubit gate infidelities, including qubit decay, heating, dephasing, carrier-envelope variations, beyond RWA effects and RF flux crosstalk. In this appendix we quantify the contribution of each error source to the infidelity via QuTiP simulations and analytical estimates (where possible). 

Simulations suggest that leakage to non-computational states is negligible, attributable to the significant detuning between relevant drive frequencies and undesired transition frequencies. We can thus safely truncate the Hilbert space to the computational subspace. 
In the following we focus on the specific example of error-budgeting two-qubit gates, with similar results holding for single-qubit gates. In the presence of a coupler-flux drive, the lab-frame Hamiltonian is 
\begin{align}
\label{eq:Htwo_q_error_budget}
H = -\frac{\omega_{a}}{2}\sigma_{z}^{a} -\frac{\omega_{b}}{2}\sigma_{z}^{b} +Af(t)\cos(\omega_{d}t + \theta_{\text{CE}})\sigma_{x}^{a}\sigma_{x}^{b},
\end{align}
where $A$ is the amplitude of the drive, $f(t)$ is the Gaussian envelope and $\theta_{\text{CE}}$ is the carrier-envelope phase. 

\begin{table*}[ht]
\caption{\label{table:error_budget} Error budget for single- and two-qubit gates. Analytic estimates of decoherence rates can be found in Ref.~\cite{Abad}. We use gate lengths $\tau=83.3, 65.1, 101.6, 257.8$ ns for the gates $\sqrt{X_a}$, $\sqrt{X_b}$, $\sbswap$ and $\siswap$, respectively. The unaccounted error is attributed to insufficient calibration and RF flux crosstalk.
}
\centering
\begin{tabularx}{\linewidth}{c *{4}{>{\centering\arraybackslash}X}}
\hline\hline
error source & $\siswap$ & $\sbswap$ & $\sqrt{X_a}$ & $\sqrt{X_b}$\\
& estimated (simulated) & estimated (simulated) & estimated (simulated) & estimated (simulated) \\
\hline \hline
decay &  $4.6\times10^{-4}$ $(4.6\times10^{-4})$ & $1.8\times10^{-4}$ ($1.8\times10^{-4}$) & $1.5\times 10^{-4}$ $(1.5\times 10^{-4})$ & $1.1\times10^{-4}$ $(1.2\times10^{-4})$\\ \hline 
heating & $4.6\times10^{-4}$ $(4.6\times10^{-4})$ & $1.8\times10^{-4}$ ($1.8\times10^{-4}$) & $1.5\times 10^{-4}$ $(1.5\times 10^{-4})$ & $1.1\times10^{-4}$ $(1.2\times10^{-4})$\\ \hline
dephasing & 
$3.0\times10^{-4}$ $(3.0\times10^{-4})$ & $1.2\times10^{-4}$ ($1.2\times10^{-4}$) & $1.0\times 10^{-4}$ $(1.0\times 10^{-4})$ & $7.5\times 10^{-5}$ $(7.5\times 10^{-5})$ \\ \hline
\hline
beyond RWA & $1.4\times10^{-4}$ ($8.4\times10^{-5}$) & $1.1\times10^{-5}$ ($1.3\times10^{-5}$) & $5.2\times10^{-5}$ ($5.6\times10^{-5}$) & $7.5\times10^{-5}$ ($5.6\times10^{-5}$) \\ \hline
carrier envelope phase & ($5.8\times10^{-5}$) & ($1.4\times10^{-5}$) & ($0.5\times10^{-5}$) & ($0.5\times10^{-5}$) \\ \hline
higher order drive terms & ($8.5\times10^{-5}$) & ($2.3\times10^{-5}$) & ($1.4\times10^{-5}$) & ($2.2\times10^{-5}$) \\
\hline \hline
estimated infidelity (decoherence) & $1.2\times10^{-3}$ & $4.8\times10^{-4}$  & $4.0\times 10^{-4}$ & $3.0\times 10^{-4}$\\
\hline
simulated infidelity (all) & $1.4\times10^{-3}$ & $5.3\times10^{-4}$ & $4.7\times 10^{-4} $ & $3.7\times 10^{-4}$\\
\hline 
measured infidelity & $2.8\times10^{-3}$ & $9\times10^{-4}$ & $6\times 10^{-4}$ & $5\times 10^{-4}$ \\
 \hline
 \hline
\end{tabularx}
\end{table*}

\subsection{Beyond the RWA}
Both qubits in our sample have Larmor periods of about 20 ns, only a slightly smaller timescale than the single- and two-qubit gates that we implement. Thus, it is possible that fidelities could begin to be limited by effects arising from counter-rotating terms. 
To obtain a semi-analytic estimate of the associated fidelity reduction, we utilize a Magnus expansion~\cite{Wilcox, Blanes2009, Magnus} following~\cite{Weiss2022}. Noting that the Hamiltonian only couples pairs of states $|00\rangle \leftrightarrow |11\rangle$ and $|01\rangle \leftrightarrow |10\rangle$, we write the Hamiltonian (J3) as a direct sum $H_{2q}(t)=H_{-}(t)\bigoplus H_{+}(t)$, where
\begin{align}
H_{\pm}(t)=-\frac{\omega_{\pm}}{2}\Sigma_{z}^{\pm}+Af(t)\cos(\omega_{d}t)\Sigma_{x}^{\pm}.
\end{align}
We have defined $\omega_{\pm}=\omega_{a}\pm\omega_{b}$ and the Pauli matrices as e.g. $\Sigma_{x}^{+} = |00\rangle\langle11| + \mathrm{H.c.}$ and $\Sigma_{x}^{-} = |01\rangle\langle10| + \mathrm{H.c.}$ For simplicity we set $\theta_{\text{CE}} = 0$, and postpone a discussion of carrier-envelope variations to Sec.~\ref{sec: error-CEP}. To isolate the effects of the drive, we move into a rotating frame, yielding
\begin{align}
\label{eq:Hpm}
H_{\pm}'(t) = Af(t)\cos(\omega_{d}t)[\cos(\omega_{\pm} t)\Sigma_{x}^{\pm}+\sin(\omega_{\pm} t)\Sigma_{y}^{\pm}].
\end{align}
Utilizing the first two terms in the Magnus expansion, the propagator is $U_{2q}(t)=\exp(\Delta_{1}^{-}[t]+\Delta_{1}^{+}[t]+\Delta_{2}^{-}[t]+\Delta_{2}^{+}[t])$, where~\cite{Wilcox, Blanes2009, Magnus}
\begin{align}
\Delta_{1,\pm}(t) &=-i\int_{0}^{t}H_{\pm}'(t')dt', \\ 
\Delta_{2,\pm}(t) &=-\frac{1}{2}\int_{0}^{t}dt_{1}\int_{0}^{t_{1}}dt_{2}[H_{\pm}'(t_{1}),H_{\pm}'(t_{2})].
\end{align}
The gate time $\tau$ is taken to be $\tau=\sqrt{2\pi}/(A\erf(\sqrt{2}))$, appropriate to obtain a $\siswap$ or $\sbswap$ in the RWA with Gaussian envelope $f(t) = \exp(-0.5 t^2/(\sigma^2))$ and width $\sigma=\tau/4$. 

We obtain for the first terms in the Magnus expansion
\begin{align}
\Delta_{1,-}(\tau)+\Delta_{1,+}(\tau) &= -i\alpha_{-}\Sigma_{x}^{-} -i\alpha_{+}\Sigma_{x}^{+} \\ \nonumber 
&= -i\frac{\alpha_{-}}{2}(\sigma_{x}^{a}\sigma_{x}^{b}+\sigma_{y}^{a}\sigma_{y}^{b}) \\ \nonumber 
&\quad-i\frac{\alpha_{+}}{2}(\sigma_{x}^{a}\sigma_{x}^{b}-\sigma_{y}^{a}\sigma_{y}^{b}),
\end{align}
where 
\begin{align}
\alpha_{\pm} = \int_{0}^{\tau}dt A f(t) \cos(\omega_{d}t)\cos(\omega_{\pm}t).
\end{align}
These terms describe the desired drive ($\alpha_{-}=\pi/4, \alpha_{+}=0$ for $\siswap$, $\alpha_{+}=\pi/4, \alpha_{-}=0$ for $\sbswap$) as well as an undesired term that causes transitions between e.g. $|00\rangle\leftrightarrow|11\rangle$ in the case of $\siswap$.
For the second-order terms we obtain
\begin{align}
\Delta_{2,-}(\tau) + \Delta_{2,+}(\tau) &= -i \beta_{-}\Sigma_{z}^{-}-i\beta_{+}\Sigma_{z}^{+} \\ \nonumber 
&= -i\frac{\beta_{+}+\beta_{-}}{2}\sigma_{z}^{a}
-i\frac{\beta_{+}-\beta_{-}}{2}\sigma_{z}^{b},
\end{align}
where 
\begin{align}
\nonumber 
\beta_{\pm} = A^2\int_{0}^{\tau}dt_{1}\int_{0}^{t_{1}}dt_{2}
f(t_{1})f(t_{2})\cos(\omega_{d}t_{1})\cos(\omega_{d}t_{2}) \\ 
\times [\cos(\omega_{\pm}t_{1})\sin(\omega_{\pm}t_{2})-\sin(\omega_{\pm}t_{1})\cos(\omega_{\pm}t_{2})].
\end{align}
These terms describe the familiar Bloch-Siegert shift, where the resonance frequencies of both qubits are shifted by the coupler drive.

As an aside, we comment that we have just derived the propagator to second order for single-qubit gates with a drive on $\sigma_{x}$ (returning to usual Pauli-matrix notation for clarity),
$U_{1q}(t) = \exp(-i\frac{1}{2}\alpha \sigma_{x} -i\frac{1}{2}\beta \sigma_{z}),$ dropping the subscripts on $\alpha$ and $\beta$ and redefining $\alpha$ and $\beta$ with an additional factor of $2$ for later convenience.

Returning to two-qubit gates, we now specialize to the case of $\siswap$ with $\omega_{d}=\omega_{a}-\omega_{b}$. A similar analysis follows for $\sbswap$ with $\omega_{d}=\omega_{a}+\omega_{b}$.
The propagator in this case is
\begin{align}
U_{2q}(\tau) = \exp\Big(
-i\frac{1}{2}\left[\frac{\pi}{4}+\delta\right][\sigma_{x}^{a}\sigma_{x}^{b}+\sigma_{y}^{a}\sigma_{y}^{b}] \\ \nonumber 
-i\frac{\alpha_{+}}{2}[\sigma_{x}^{a}\sigma_{x}^{b}-\sigma_{y}^{a}\sigma_{y}^{b}] \\ \nonumber 
-i\frac{\beta_{+}+\beta_{-}}{2}\sigma_{z}^{a}
-i\frac{\beta_{+}-\beta_{-}}{2}\sigma_{z}^{b}\Big),
\end{align}
where we have isolated the deviation of $\alpha_{-}$ from the ideal value of $\pi/4$, $\alpha_{-}=\frac{\pi}{4}+\delta$.
We may now read off the Bloch-Siegert shifts $\omega_{\mathrm{BS},\mu}$, given by $\omega_{\mathrm{BS},a}=(\beta_{+}+\beta_{-})/\tau$ and $\omega_{\mathrm{BS},b}=(\beta_{+}-\beta_{-})/\tau$.
With the propagator in hand, we may now compute the gate fidelity using the standard formula 
~\cite{Pedersen2007}
\begin{align}
\label{eq:F}
F=\frac{\Tr[U^{\dagger}(\tau)U(\tau)]+\Big|\Tr[U_{\mathrm{T}}^{\dagger}U(\tau)]\Big|^2}{d(d+1)},
\end{align}
where $d$ is the dimension of the Hilbert space and $U_{\rm T}$ is the target unitary. Expanding $\Tr[U_{\mathrm{T}}^{\dagger}U(\tau)]$ about 
$(\delta,\alpha_{+},\beta_{-},\beta_{+})=(0,0,0,0)$ and retaining only leading-order terms, we obtain
\begin{align}
F = 1-\frac{2}{5}(\alpha_{+}^{2}+\delta^2+\beta_{+}^{2}+\frac{8}{\pi^2}\beta_{-}^{2}),
\end{align}
in the case of $\siswap$ and 
\begin{align}
F = 1-\frac{2}{5}(\alpha_{-}^{2}+\delta^2+\beta_{-}^{2}+\frac{8}{\pi^2}\beta_{+}^{2}),
\end{align}
in the case of $\sbswap$ (with $\delta$ now defined as $\alpha_{+}=\frac{\pi}{4}+\delta$).
Evaluating the integrals $\alpha_{\pm}, \beta_{\pm}$ numerically, we obtain $1-F$=$1.4\times10^{-4}, 1.1\times10^{-5}$, in the cases of $\siswap, \sbswap$, respectively. For the Bloch-Siegert shifts, we obtain 
$|\omega_{\mathrm{BS}, a}|/2\pi=9$ kHz, $|\omega_{\mathrm{BS}, b}|/2\pi=20$ kHz in the case of $\siswap$ and $|\omega_{\mathrm{BS}, a}|/2\pi=11$ kHz, $|\omega_{\mathrm{BS}, b}|/2\pi=2$ kHz in the case of $\sbswap$.

We compare the fidelity results with estimates obtained by numerical simulation of Eq.~\eqref{eq:Hpm} using QuTiP~\cite{qutip2}. These simulations yield
infidelity contributions of $8.4\times10^{-5}$, $1.3\times10^{-5}$ for $\siswap$ and $\sbswap$, respectively. In both cases, the analytically-estimated values are within about a factor of two of the numerically obtained infidelities.

In the case of single-qubit gates with $U_{\rm T}=\sqrt{\textit{X}_{\mu}}$, we obtain for the fidelity to leading order 
\begin{align}
F = 1 - \frac{4\beta^2}{3\pi^2}-\frac{\delta^2}{6},
\end{align}
where $\delta=\alpha-\frac{\pi}{2}$. We obtain infidelity estimates of $7.8\times10^{-5}, 7.5\times10^{-5}$ for $\sqrt{\textit{X}_{a}}$ and $\sqrt{\textit{X}_{b}}$, respectively. Performing numerical simulations (with the single-qubit version of Eq.~\eqref{eq:Hpm}) we obtain infidelities of $5.6\times10^{-5}$ in both cases. These results are slightly lower than the analytic estimates, but of the same order of magnitude.

\subsection{\label{sec: error-CEP}Carrier envelope variations}
The phase $\theta_{\text{CE}}$ is constantly updated during the experiment. As the logical states are defined in the rotating frame, the phase of the pulse should be matched with the dynamical phase difference of the states being swapped. Besides, we make use of the phase $\theta_{\text{CE}}$ to realize virtual $\textit{Z}$ gates. 
Because the pulse length is only 4 (11) drive periods in the case of the $\siswap$ ($\sbswap$) gate, the energy carried by the pulse is generally a function of $\theta_{\text{CE}}$ (to be contrasted with the more standard case when the pulse length is much larger than the drive period, and there is no such dependence on $\theta_{\text{CE}}$). Additionally, the effects of the counter-rotating terms can depend on $\theta_{\text{CE}}$. To quantify the contribution of this source of error,
we numerically sweep the phase of the pulse and calculate the gate infidelity as a function of $\theta_{\text{CE}}$. The average of the increase in infidelity due to phase variation is what is taken as the contribution from carrier envelope phase in Table \ref{table:error_budget}.

\subsection{Higher-order drive terms}
As discussed in Ref.~\cite{Weiss2022}, the form of the drive operator in Eq.~\eqref{eq:Htwo_q_error_budget} is an approximation. In reality, the drive operator associated with $\Phi_{\text{ext},c}$ is not perfectly $XX$ but includes also nonzero $IX$, $XI$ and other components. Similarly for the drive operators associated with $\Phi_{\text{ext},a}$, $\Phi_{\text{ext},b}$, the drive operators contain nonzero $ZX$, $XZ$ and other terms. Including these additional drive terms in the simulations contributes infidelities on the order of $10^{-5}$, see Table ~\ref{table:error_budget}.

\subsection{Decoherence}

The gate fidelity is reduced due to decay, heating and dephasing effects. In the following we specialize to the case of the $\siswap$ gate, taking $\omega_{d}=|\omega_{b}-\omega_{a}|$. The discussion follows similarly for the case of $\sbswap$, with $\omega_{d}=\omega_{b}+\omega_{a}$. The Hamiltonian after performing the RWA is
\begin{align}
H_{\rm RWA}' = \frac{A}{4}(\sigma_{x}^{a}\sigma_{x}^{b}+\sigma_{y}^{a}\sigma_{y}^{b}).
\end{align}
We perform numerical simulations of the Lindblad master equation
\begin{align}
	\frac{d\rho(t)}{dt}&=-i[H_{\rm RWA}', \rho(t)] +\sum_{k}\Gamma_{k}\mathcal{D}(L_{k})\rho(t)
\end{align}
where 
\begin{align*}
	\mathcal{D}(L)\rho = L\rho L^{\dagger}-\frac{1}{2}\{L^{\dagger}L,\rho\},
\end{align*}
is the dissipator and $\rho(t)$ is the system density matrix.
We include collapse operators $\sigma_{-}^{\mu}, \sigma_{+}^{\mu}, \sigma_{z}^{\mu}$ and associated decay rates $\Gamma_{1}/2, \Gamma_{1}/2, \Gamma_{\phi}/2$, respectively ($\mu=\{a,b\}$). The factors of $1/2$ in the decay and heating rates are due to the equal contributions of decay and heating to $T_{1}$, while the factor of $1/2$ in the dephasing rate ensures that coherences decay at the appropriate rate ~\cite{Abad}. By turning on each decoherence channel one at a time, we numerically estimate the contribution of each, see Table ~\ref{table:error_budget}. We then proceed to turn on all decoherence channels to estimate the overall fidelity reduction due to decoherence. For both the $\sbswap$ and $\siswap$ gates, decoherence makes up about half of the measured infidelity.

Analytical estimates of the first-order contribution of decoherence to the infidelity are given in Ref.~\cite{Abad}, which we reproduce here for completeness
\begin{align}
F\approx 1-\frac{d}{2(d+1)}\tau\sum_{\mu}(\Gamma_{1}^{\mu}+\Gamma_{\phi}^{\mu}),
\end{align}
where $\tau$ is the duration of the gate and $d$ is the dimension of the Hilbert space. Infidelity estimates using this formula for single- and two-qubit gates can be found in Table ~\ref{table:error_budget}. We find excellent agreement between the numerical results and analytical formulas.

\subsection{RF flux crosstalk}

As discussed in Appendix~\ref{app:acflux}, flux crosstalk can contribute to dynamic frequency shifts and unwanted entanglement. Defining the maximum fluxes due to crosstalk $\xi_{\mu}$ through the qubit loops,
the qubit drive amplitudes are $\Omega_{\mathrm{ct}, \mu} = \xi_\mu E_{L\mu}\langle0|\phi_{\mu}|1\rangle$, $\mu = a, b$. 
To estimate the infidelity associated with such crosstalk, we substitute Eq.~\eqref{eq:H_cross} for Eq.~\eqref{eq:Htwo_q_error_budget}, including the Gaussian envelope and transforming into the rotating frame with the qubit frequencies. For $\xi_\mu/2\pi=10^{-4}$, we find numerically that the infidelity is $1.2\times10^{-6}, 1.2\times10^{-5}$ for $\sbswap$ and $\siswap$, respectively, below the levels we are sensitive to. However, if crosstalk rises to the level of $\xi_\mu/2\pi=10^{-3}$, the infidelities rise to
$1.1\times10^{-3}, 0.069$ for $\sbswap$ and $\siswap$, respectively. These infidelities would be the leading source of error, necessitating the careful flux-crosstalk cancellation we undertook in our experiment. 

\subsection{Overall error budget}
With all of the error channels included, the infidelities of single qubit gates, $\siswap$, and $\sbswap$ gates are reduced to $99.97\%$, $99.87\%$, and $99.95\%$ respectively. Our numerical simulations suggest that gate fidelity is predominantly constrained by decoherence. The remaining error is attributed to insufficient calibration and RF flux crosstalk.

\bibliography{thebibliography}

\end{document}